# Reversal in time order of interactive events: Collision of inclined rods

**Chandru Iyer[1] and G. M. Prabhu[2]**

[1]Techink Industries, C-42, phase-II, Noida, India, Contact E-mail: chandru_i@yahoo.com
[2]Department of Computer Science, Iowa State University, Ames, IA, USA
Contact E-mail: prabhu@cs.iastate.edu

**Abstract**
In the rod and hole paradox as described by Rindler (1961 *Am. J. Phys.* **29** 365-6), a rigid rod moves at high speed over a table towards a hole of the same size. Observations from the inertial frames of the rod and slot are widely different. Rindler explains these differences by the concept of differing perceptions in rigidity. Grøn and Johannesen (1993 *Eur. J. Phys.* **14** 97-100) confirmed this aspect by computer simulation where the shapes of the rods are different as observed from the co-moving frames of the rod and slot. Lintel and Gruber (2005 *Eur. J. Phys.* **26** 19-23) presented an approach based on retardation due to speed of stress propagation. In this paper we consider the situation when two parallel rods collide while approaching each other along a line at an inclination with their axis. The collisions of the top and bottom ends are reversed in time order as observed from the two co-moving frames. This result is explained by the concept of 'extended present' derived from the principle of relativity of simultaneity.

**Introduction**

Measuring times and time intervals involves the concept of simultaneity. When a person says he awoke at six o'clock he means that two events, his awakening and the arrival of the hour hand of his clock at the number six, occurred simultaneously. The fundamental problem in measuring time intervals is that in general two events that appear simultaneous in one frame of reference do not appear simultaneous in a second frame which is moving relative to the first, even if both are inertial frames.

A basic problem is this: When an event occurs at point *(x, y, z)* at time *t,* as observed in a frame of reference S, what are the coordinates *( x′, y′, z′ )* and time *t′* of the event as observed in a second frame S′ moving relative to S with constant velocity *v* along the *x*-direction?

Einstein [1] showed that the Lorentz transformation equations describe the relation between (x, y, z, t) and (x', y', z', t') under the principles of equivalence of inertial frames and constancy of the speed of light. The Lorentz transformation equations, according to the principle of relativity, describe differences in observation depending on motional speed. If the length of an object at rest is $L_0$, then when it moves relative to a coordinate system with a speed *v* it is expected to appear contracted from $L_0$ to L according to the relationship:

$$L = L_0 / \gamma$$

where $\gamma = \frac{1}{\sqrt{1 - v^2/c^2}}$ and $c$ is the speed of light in vacuum. In a co-moving coordinate system, the length $L'$ will still appear to be equal to $L_0$.

Mutual length contraction has always confounded students of special relativity (SR), leading to two sets of paradoxes. One is the train-tunnel paradox and the other is the rod-slot paradox. In the former, the question of interest is whether the train is smaller than the tunnel or the tunnel is smaller than the train. In the latter, the question of whether the rod falls into the slot or not has been widely discussed.

While Rindler [2] advanced the idea of differing perceptions of rigidity to explain the observations by the rod and the slot, Shaw [3] felt that as the rod starts to 'fall,' it experiences accelerated motion. Under those conditions it will look curved from one frame and straight from another. Shaw [3] further opined that the paradox can be taken out of the purview of gravity and accelerated motion by considering a different problem where there is no gravity and therefore more strictly under the special theory of relativity. He proposed a slot and a rod approaching each other (as seen from a base inertial frame F). The rod has a large velocity along the x-axis with respect to F, while the slot has a small velocity in the z-axis (with respect to F) moving in such a way that their centers coincide at an instant $t = 0$ in frame F. He showed that while the slot perceives that the rod fell into it with their lengths fully aligned, the rod perceives that it went into the slot in an inclined way and thus could go into the slot even though it (the rod) was longer than the slot.

Grøn and Johannesen [4] simulated the falling of the rod into the slot by developing a computer program which numerically transforms the coordinates from one frame to another by the Lorentz transformation equations and displays the view from the co-moving frames of the rod and the slot by computer graphics. The graphics vividly show that the shape and inclination of the rod is very different as observed from the two inertial frames. This confirms Rindler's conjecture that the perceptions of 'rigidity of the rod' may vary as seen by the two frames. The question of whether the nature of a physical effect depends upon the frame of reference from which the said effect is observed is addressed in [4] by emphasizing on the need for a relativistic theory of elasticity. The authors note that while a breakage in the rod is a physical effect, the bending of the rod is not a physical effect, as different reference frames perceive the extent of this bending differently. As yet, to our knowledge, no detailed formulation on a relativistic theory of rigidity/elasticity has been proposed in the literature.

The reason why an object looks different in length and in shape from different inertial frames can be understood when we recognize that in SR, an object as an entity is not recognized as such but only as a set of events; when these events transform between frames, they take different dimensions and shapes. An event (a collection of which makes up an object), does not 'belong' to an inertial frame or an object.

The rod-slot paradox was recently re-visited in [5]. According to [5] the car and hole paradox described in [6] and the rod and hole paradox described in [2 and 3], do not give rise to any contradiction when explained on the basis of stress propagation – that is, the difference in observed speed of stress propagation compensates exactly for the differences in observed length.

In this paper we describe a paradox called the "Collision of inclined rods" and discuss the reversal in time order of events as observed from two inertial frames. Our presentation has two important pedagogical contributions: (a) we describe a situation in 2-d space, and (b) the rod and slot paradox becomes a special case of the collision of inclined rods paradox. We show how the concept of 'extended present,' which is a corollary of the principle of relativity of simultaneity can be used to satisfactorily resolve this paradox.

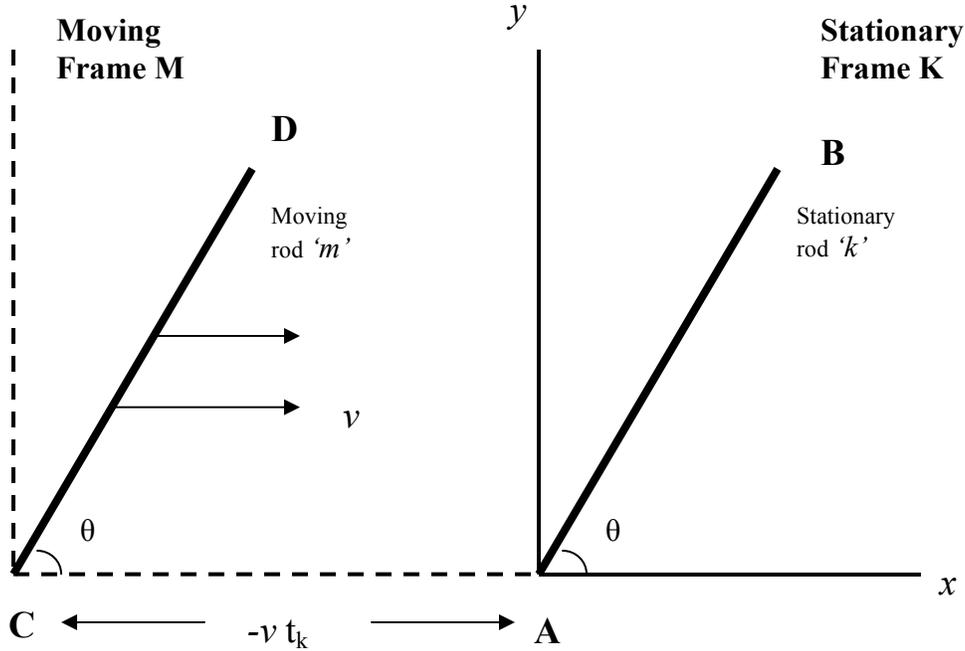

**Figure 1.** Identical rods in frames K and M. Shown above is an instant $-t_k$ of frame K.
$A = (0, 0, -t_k)$ ; $B = (a, h, -t_k)$ ; $C = (-v\, t_k, 0, -t_k)$ ; $D = ((a/\gamma) - v\, t_k, h, -t_k)$

**1. Inclined rods**

Consider 2 two-dimensional frames K and M as shown in Figure 1. Both frames have identical rods (identical in their respective rest frames), placed as indicated at an angle θ with the x-axis. The frame K is stationary with inclined rod '$k$' and the frame M is moving with inclined rod '$m$.' The angle that rod '$m$' (CD) makes with the x-axis is also θ in frame M, but it appears to be different as seen from K due to the Lorentz contraction of the projection of CD on the x-axis. There is no such contraction for the projection on the y-axis.

The relative velocity of M with respect to K is $v$ along the x-axis, and as M and K approach each other the two rods will collide. An observer in frame K sees the rod in frame M approaching with the top end tilted towards him. An observer in frame M sees the rod in frame K approaching with the top end inclined away from him. The bottom ends of the rods are some distance apart on the x-axis. The lengths of the rods as seen by the respective co-moving frames are the same.

According to Newtonian mechanics, the two rods will collide in such a way that their entire lengths come into contact with each other at one *single* instant.

We show in the next section that this is not the case in relativistic mechanics. As seen from frame M, the top ends of the rods collide first and the bottom ends collide later. As seen from frame K, the bottom ends of the rods collide first and the top ends collide later. This paradox is not explained by stress propagation because the direction of propagation of the stress in the rods is reversed in both frames.

Thus, as explained below, this paradox presents a real contradiction.

**2. Collision of inclined rods**

The coordinates of the top end of rod '*k*' in frame K are x = *a* and y = *h* for all times t, since rod '*k*' is stationary in frame K. Similarly, the coordinates of the top end of rod '*m*' in frame M are x′ = *a* and y′ = *h* for all times t′ since rod '*m*' is stationary in frame M. We define the origin of space and time at the instant when the bottom ends of the rods collide as x = 0, y = 0, t = 0, or as the triple (0 , 0 , 0) for this event in both frames K and M. Thus the point marked C (see figure 1) is located at x = –v $t_k$ with respect to frame K and after a time $t_k$ elapses, C will travel a distance (v $t_k$ ) and will reach point A and collide.

At the instant –$t_k$, the coordinates of the points A, B, C, and D as seen from frame K are:

A = (0 , 0 , –$t_k$ ) ; B = (*a* , *h* , –$t_k$ ) ; C = (– v $t_k$ , 0 , –$t_k$ ) ; D = ( ((*a* / γ) – v $t_k$ ) , *h* , –$t_k$ )

where $\gamma = 1/\sqrt{1-(v^2/c^2)}$  and *c* is the speed of light in vacuum.

From the Lorentz transformations, the coordinates of point C in frame M are as follows.

$C_x$ = [– v $t_k$  –  v ( – $t_k$) ] γ  =  0

$C_y$ = 0

$C_t$ = [–$t_k$  –  (– v $t_k$ ) v / $c^2$ ] γ  =  – $t_k$ / γ

So the point C in frame M = ( 0 , 0 , – $t_k$ / γ ).

Similarly, the coordinates of D in frame M turn out to be ( *a* , *h* , – {$t_k$ / γ  + a v / $c^2$ }). Notice that the time coordinates of C and D in M do not match because Figure 1 is an instant in frame K but not an instant in frame M.

By setting $t_k$ = 0, we get the collision of A and C in both frames K and M at (0, 0, 0).

In order for a collision of B and D to occur in frame K, all space and time coordinates of B and D in frame K must be equated.

(*a* / γ) – v $t_k$  = *a*
Thus, $t_k$ = (*a* / v ) ((1 / γ)  – 1).

Since the instant we are considering in frame K is –$t_k$, the time of collision is (*a* / v ) (1 –  (1 / γ) ), which is positive in frame K.

When we record the observation in frame M, we need only change v to – v.  The time of collision of A and C remains unchanged and occurs at time t′ = 0. The time of collision of B and D in frame M is (*a* / v ) ((1 / γ)  – 1), which is negative in frame M.

## Snapshots in Frame K

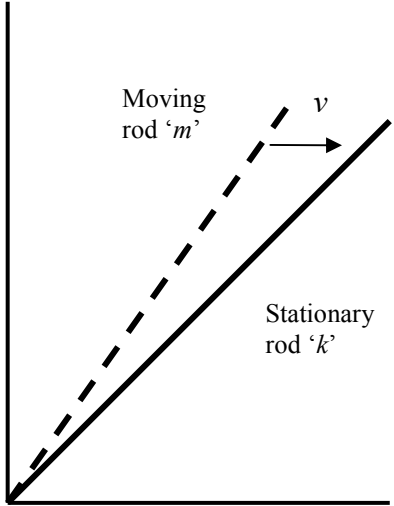

**Figure 2 (i).** Instant t = 0 in K
Bottom collision happening now
Top collision yet to happen

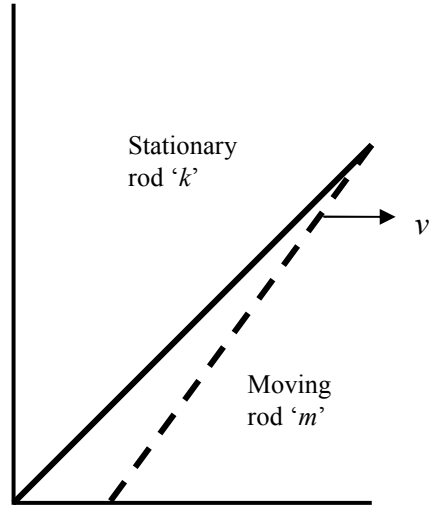

**Figure 2 (ii).** Instant t = ($a/v$) (1 − (1 / $\gamma$) ) in K
Top collision happening now
Bottom collision already happened

## Snapshots in Frame M

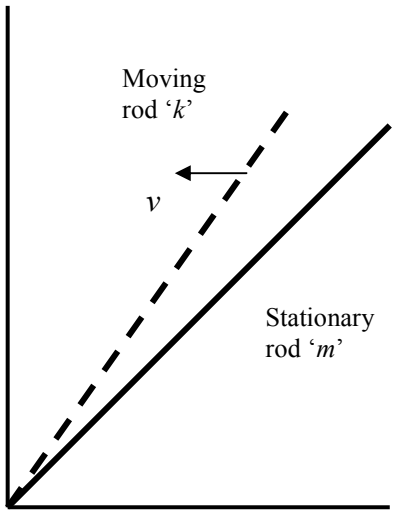

**Figure 3 (i).** Instant t′ = 0 in M
Bottom collision happening now
Top collision already happened

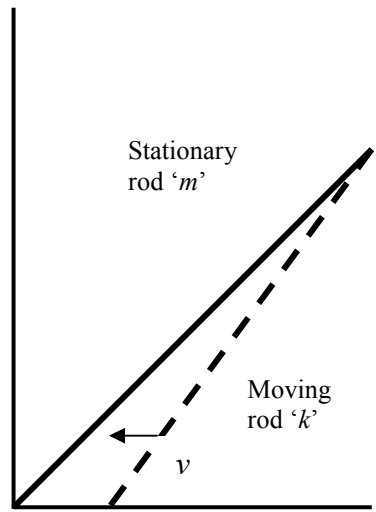

**Figure 3 (ii).** Instant t′ = ($a/v$) ((1/$\gamma$) − 1) in M
Top collision happening now
Bottom collision yet to happen

If one analyzes the above figures carefully one sees that when we invert figure 2 (i) we get figure 3 (ii) and vice versa. Similar behavior is exhibited between figures 2 (ii) and 3 (i). Since there is no 'top' or 'bottom' in actuality we find that the system exhibits symmetry as one would expect.

Which of the ends collided first in a given frame is determined by the following principle: For an observer who is located at the center of any one of the 'stationary' rods, the end of his rod that is tilted towards the approaching 'moving' rod will collide first.

Summarizing, for frame K, the collision of the bottom of the rods occurs at time t = 0, and the collision of the top of the rods occurs at a later instant (t > 0). For Frame M, the collision of the bottom of the rods occurs at time t′ = 0, and the collision of the top of the rods occurs at an earlier instant (t′ < 0). We note that the sum of the times at which the top ends of the rod collide in both frames adds up to zero. This implies that one of them is negative and the other is positive. Both cannot be zero unless $v = 0$.

It is clear that when the bottom collision of the rods occurred at time t = 0 in both frames, the top collision occurred prior to that in frame M (in the past, because t′ < 0) and occurred later in frame K (in the future, because t > 0). As seen from frame M, the stress propagates from bottom to top in both the rods 'k' and 'm'; however, as seen from frame K, the stress propagates from top to bottom in both the rods 'k' and 'm.'

In the Newtonian case, one can use the Galilean transformation x′ = x – vt and conclude that the rod 'm' collides along its entire length with rod 'k' at one single instant in both frames at t = 0.

The collision of the inclined rods paradox is a generalized version of the familiar train-tunnel and rod-slot paradoxes. If we let $h = 0$ and make one rod a hollow cylinder, we get the train-tunnel paradox, which is closely related to the rod-slot paradox.

**3. Discussion**

This paradox may be resolved by invoking the concept of "extended present" [7], which is an inevitable consequence of the principle of relativity of simultaneity. In the Galilean model, given an event E, the set of events S simultaneous with E, can be determined without reference to any characteristics of a specific inertial frame. In special relativity, the set S depends on the characteristics of each reference frame observing the event E. However, in SR, the set of extended present events $P_E$ of a given event E can be defined without reference to any characteristics of a specific inertial frame. The set $P_E$ is the union of all possible sets S (for event E) associated with all inertial frames. Every event in $P_E$ is simultaneous with E in some inertial frame. To illustrate, for an inertial frame K, if event E has coordinates $x = 0; t = 0$ and another event G has coordinates $(a, t)$, then the necessary and sufficient condition that G belongs to $P_E$ is that the absolute value of $a/t$ is greater than $c$; furthermore another inertial frame M moving at a velocity $c^2t/a$ with respect to K will see E and G as simultaneous. Thus we view the concept of $P_E$ as an extension of relativity of simultaneity.

For any observer (in a given inertial frame), $P_E$ contains 'future' as well as 'past' events over which event E has no influence or cannot be influenced by and this means that there can be no causal link between E and any event in $P_E$. Formally, no signal traveling at or below the speed of light, $c$, can link E with any event in $P_E$.

Our paradox is resolved by observing that for both frames K and M, the top collision lies in the extended present of the bottom collision and vice versa. The reversal in time order has no physical significance, as there can be no causal link between the two events.

When we consider a frame F that is moving at a velocity $+u$ with respect to K and $-u$ with respect to M, where $u = v / (1 + (1/\gamma))$, we find that the two collisions are simultaneous as seen by F. From relativity of simultaneity we observe that the frames K and M perceive these two events as non-simultaneous, but within the scope of an 'extended present.'

Although, these concepts are well known, the paradox presented here is significant in that the two events for which the reversal in time order occurs are linked by a continuous object (a rod) in both the frames. But an object, such as a rod, is recognized only as a collection of events and has no independent physical sanctity. In the present case, there can be no causal link between the two events unless a disturbance travels at a speed greater than $c$, which is precluded under SR.

**References**

[1]  Einstein A *Relativity, The Special and General Theory*, Authorized Translation by Robert W. Lawson, Three Rivers Press, New York, 1961
[2]  Rindler W 1961 Length contraction paradox *Am. J. Phys.* **29** 365-6
[3]  Shaw R 1962 Length contraction paradox *Am. J. Phys.* **30** 72
[4]  Grøn Ø and Johannesen S 1993 Computer simulation of Rindler's length contraction paradox *Eur. J. Phys.* **14** 97-100
[5]  Lintel H and Gruber C 2005 The rod and hole paradox re-examined *Eur. J. Phys.* **26** 19-23
[6]  Bolstein A 2004 *Physical Proof of Einstein's Relativity Failure* Online at http://www.wbabin.net/physics/ppef.htm
[7]  Resnick R *Introduction to Special Relativity,* John Wiley and Sons, 1968 Appendix A3 – The Time Order and Space Separation of Events